\documentclass[pdflatex,sn-basic]{sn-jnl}

\usepackage{amsmath}
\usepackage{amssymb}
\usepackage{float}
\usepackage{booktabs}
\usepackage{multirow}
\usepackage{tabularx}
\theoremstyle{thmstyleone}%

\theoremstyle{thmstyletwo}%

\theoremstyle{thmstylethree}%

\begin{document}

\title[Comparing Visual Motion Patterns for Pedestrian Guidance]{Comparing Visual Motion Patterns for\\Pedestrian Guidance with Multi-Display Signage: Two~Field~Studies}

\author*[1,2]{\fnm{Yuri} \sur{Mikawa}}\email{yuri\_mikawa@ipc.i.u-tokyo.ac.jp}
\author[3]{\fnm{Taiki} \sur{Fukiage}}
\author[3]{\fnm{Yuki} \sur{Kubota}}
\author[3]{\fnm{Takumi} \sur{Yokosaka}}
\author[3]{\fnm{Kazushi} \sur{Maruya}}

\affil*[1]{\orgname{The University of Tokyo}, \orgaddress{\city{Tokyo}, \country{Japan}}}
\affil[2]{\orgname{JST PRESTO}, \orgaddress{\city{Tokyo}, \country{Japan}}}
\affil[3]{\orgdiv{Communication Science Laboratories}, \orgname{NTT, Inc.}, \orgaddress{\state{Kanagawa}, \country{Japan}}}

\abstract{
Public multi-display signage (MDS) has the potential to provide subtle pedestrian guidance through visual motion while retaining its conventional role in advertising and information delivery. However, it remains unclear how different visual motion patterns relate to pedestrian spatial distributions and how these relationships vary across deployment settings. We conducted two exploratory field studies in a museum, involving 18,402 pedestrian passages over 25 experimental days, to examine lateral positions in relation to stripe-motion contrast and coverage patterns. In Test Case 1 (TC1), involving predominantly one-way pedestrian flow, partial stripe presentation was associated with more leftward lateral positions than wider-coverage patterns in some comparisons, and lower contrast in full-screen stripes was not associated with a detectable reduction in the observed lateral bias. In Test Case 2 (TC2), conducted in a different spatial configuration permitting bidirectional walking, lateral positions differed by stripe-motion direction during one-way flow, but coverage effects on lateral position and stripe–advertisement differences in lateral position or passing-side choice were not significant. In both test cases, partial stripes received lower self-reported sickness ratings than full-screen stripes. These findings suggest that lower-contrast and reduced-motion-coverage designs hold practical potential for pedestrian guidance, with spatially restricted cues potentially allowing such guidance to coexist with existing advertising while reducing discomfort. At the same time, the contrasting patterns in TC1 and TC2 suggest that spatial context may shape how pedestrians respond to stripe coverage, highlighting the context-dependent suitability of visual-motion-based guidance and the need for further investigation.}

\keywords{
Vection; Multi-display signage; Public displays; Pedestrian behavior; Field study
}

\maketitle

\section{Introduction}

Public spaces commonly use textual or symbolic signage to guide pedestrians. Such signage requires attention to and interpretation of instructions, and its proliferation may increase cognitive load and visual stress \citep{zhang2026wayfinding, gao2024systematic}. Multi-display signage (MDS), already used for advertising and information delivery, can present dynamic, spatially distributed visual stimuli within otherwise largely static environments. This offers a way to subtly influence pedestrian behavior without explicit directional instructions \citep{mikawa-APMAR2025}. Previous laboratory and museum studies have reported changes in pedestrian positions associated with visual motion, including biases toward the direction of full-screen stripe motion \citep{furukawa2011vection,mikawa-vection2026}.

Existing studies have tested a limited range of stimuli, such as full-screen moving stripes and random dots in black and white, under laboratory conditions or in a single spatial configuration \citep{furukawa2011vection,mikawa-vection2026}. These studies showed that the motion of such simple geometric patterns can modulate pedestrian trajectories in both laboratory and real-world settings. It remains, however, unclear how stimulus contrast, coverage pattern, and spatial layout relate to pedestrian distributions when viewing distances, walking paths, and display exposure vary naturally. Examining motion with lower contrast or reduced coverage is therefore relevant to guidance designs that accommodate advertising and integrate with their surroundings. %

We examine the potential and limitations of visual-motion-based guidance in everyday pedestrian flows. Motivated by reports of positional biases in the direction of stimulus motion \citep{mikawa-vection2026}, we conducted two exploratory field studies to examine how pedestrian responses vary with stimulus parameters, spatial configuration, and walking direction. Both studies took place in the free-admission area of the National Museum of Emerging Science and Innovation (Miraikan) in Japan, allowing observation of naturally occurring pedestrian flows and viewing conditions that are difficult to reproduce in laboratories.

Test Case 1 (TC1) examined how stripe contrast, coverage pattern, and display spacing relate to pedestrian spatial distributions under predominantly one-way flow. Test Case 2 (TC2) examined motion direction, coverage pattern, and passing-side choice in a different spatial configuration that permits bidirectional walking. The two test cases involved 3,982 pedestrians over nine experimental days and 14,420 pedestrians over sixteen experimental days, respectively. In both studies, we combined LiDAR measurements of aggregate pedestrian distributions with questionnaire ratings of stimulus awareness, perceived influence on walking, and discomfort to characterize behavioral and subjective responses to visual motion. Drawing on an integrated discussion of both test cases, we consider the potential and limitations of using different visual-motion patterns to modulate pedestrian behavior in public spaces.

\section{Related work}

\subsection{Vection-based behavioral guidance}

Visual motion and optic flow have long been used to influence locomotion and spatial behavior. Optic flow contributes directly to the control of human walking, and manipulating the visual scene can alter a pedestrian’s heading or physical walking trajectory \citep{warren2001optic,rothacher2018visual}. Building on this principle, several studies have investigated vection-based guidance. Tanikawa et al. presented optic flow in the peripheral visual field using a head-mounted vection display and demonstrated the possibility of guiding users without explicit directional instructions \citep{tanikawa2012reflex}. Similarly, Watanabe et al. reported that peripheral optic-flow stimulation induced body sway and lateral changes in walking direction \citep{watanabe2013pedestrian}. These studies suggest that visually induced self-motion can be used not only to alter subjective perception but also to modulate actual locomotor behavior.

Simple moving patterns have been widely used to study visually induced self-motion, or vection \citep{dichgans1978visual,howard1994vection,brandt1973differential,andersen1985induced,palmisano1998stimulus,tamada2015roles,palmisano2015future}. As reviewed by \citet{seno2009}, several stimulus properties have been associated with stronger vection, including low spatial frequencies of approximately 0.045–0.286 cycles/deg (3.5–22.2$^\circ$/cycle) \citep{yu2010perception}, low temporal frequencies of 2–7 Hz \citep{sauvan1995spatiotemporal}, slow angular velocities of 5–30$^\circ$/s \citep{sauvan1993properties,riecke2011compelling,wang2021visually}, high contrast \citep{yu2010perception,guo2021effects}, peripheral stimulation \citep{brandt1973differential,palmisano1998stimulus}, and figure–ground organization \citep{kitazaki2003attentional}. These findings provide a basis for selecting stimulus parameters.

Laboratory studies have also examined how stimulus area and motion strength relate to bodily responses. Increasing the area of moving visual stimulation has been reported to strengthen subjective vection \citep{nakamura1998stimulus}. Lestienne et al. found that visually induced postural displacement increased with the velocity and density of a moving scene and also depended on its displayed area \citep{lestienne1977postural}. During treadmill walking, optic-flow patterns presented in either central or peripheral vision produced direction-specific compensatory body sway \citep{bardy1999role}. More recently, Horiuchi et al. found that increasing the amplitude of visual motion increased both vection ratings and anterior–posterior center-of-pressure displacement \citep{horiuchi2021postural}. These findings motivate the expectation that stronger visual-motion stimulation should produce greater postural or locomotor modulation. However, subjective vection strength and bodily responses are not necessarily directly coupled: Horiuchi et al. found no significant correlation between individual vection measures and postural sway within each stimulus condition.

Vection also depends on figure–ground organization, rather than motion area alone. Motion perceived as part of the background tends to induce stronger vection than motion perceived as an object \citep{seno2009object}, and the addition of static visual components can either facilitate or inhibit vection depending on their relationship to the moving component \citep{seno2010static}. These findings suggest that reducing or restructuring the moving region does not necessarily lead to a proportional reduction in vection. Together, these findings motivate comparing Full, Circle, and Partial designs that differ in the area and arrangement of moving stripes and advertising content.

\subsection{In-the-wild deployments of public displays}

Public displays have been extensively studied through deployments in real-world public environments. CityWall, a large multi-touch display installed in central Helsinki, attracted 1,199 users over an eight-day deployment and revealed interaction patterns specific to public settings, including parallel interaction, collaboration, crowding, and negotiation among users \citep{peltonen2008mine}. Looking Glass investigated how passers-by recognize the interactivity of a shop-window display by presenting visual feedback responsive to their incidental body movements \citep{mueller2012looking}. Field observations of existing public displays have further shown that simply installing a display does not guarantee attention. Rather, attention to public displays depends on multiple situational and contextual factors, including their position, content, visual dynamics, and relevance to the surrounding environment \citep{huang2008overcoming, muller2009display, parker2018public}. 
Beyond display content, the physical configuration of multiple displays can itself shape how people notice and behave around them. For example, ten Koppel et al. demonstrated that different configurations of chained public displays affected the behavior of actors, audiences, and passers-by \citep{tenkoppel2012chained}.

Public displays have also been applied to navigation and the experience of pedestrian movement. Coenen et al. deployed synchronized wayfinding information across multiple consecutively positioned displays in a metropolitan railway station, demonstrating how directional information could be integrated with commercial content on the same displays \citep{coenen2016synchronized}. Monastero and McGookin developed Traces, a reactive floor projection that visualized pedestrians' walking trajectories in a university-building lobby. Their six-week field deployment examined how the in-situ presentation of these trajectories supported awareness of other people and their activities \citep{monastero2018traces}. These studies illustrate the potential of public-display systems to become part of the spatial and navigational experience rather than serving solely as isolated information or advertising media.

These deployments primarily address attention, explicit interaction, wayfinding, and social awareness. Building on preliminary work on visual-motion guidance \citep{mikawa-vection2026}, our study examines how coordinated visual motion across an MDS relates to the spatial distribution of passers-by whose walking routes were not prescribed, while retaining compatibility with advertising.

Most visual-motion guidance studies have used controlled laboratory settings \citep{furukawa2011vection,sakamoto2019guided}, with limited evidence from natural pedestrian flows \citep{mikawa-vection2026}. Awareness of an experimental setting can influence behavior \citep{orne1962social,nichols2008good}, and museum research has documented differences between laboratory and in-the-wild behavior \citep{hornecker2012wild}. Field deployments therefore complement laboratory studies by examining responses under everyday viewing and walking conditions.

\section{Experimental design and general methods}\label{sec:general-methods}
We conducted field experiments in two test cases (TC1 and TC2) to examine how the relationship between visual stimuli and pedestrian flow varies across spatial settings. TC1 examined how stimulus design and display spacing relate to pedestrian lateral positions under predominantly one-way flow. TC2 examined stimulus direction, walking direction, and passing-side choice in a different spatial configuration permitting bidirectional walking.

This section describes the visual stimuli, spatial configurations, apparatus, participant procedures, and questionnaire used in the two test cases. Case-specific details of the spatial configurations, experimental conditions, participant counts, and analyses are described in the corresponding test-case sections.

\subsection{Visual stimulus design}\label{sec:visual-stimuli-design}

\begin{figure}[t]
    \centering
    \includegraphics[width=\linewidth]{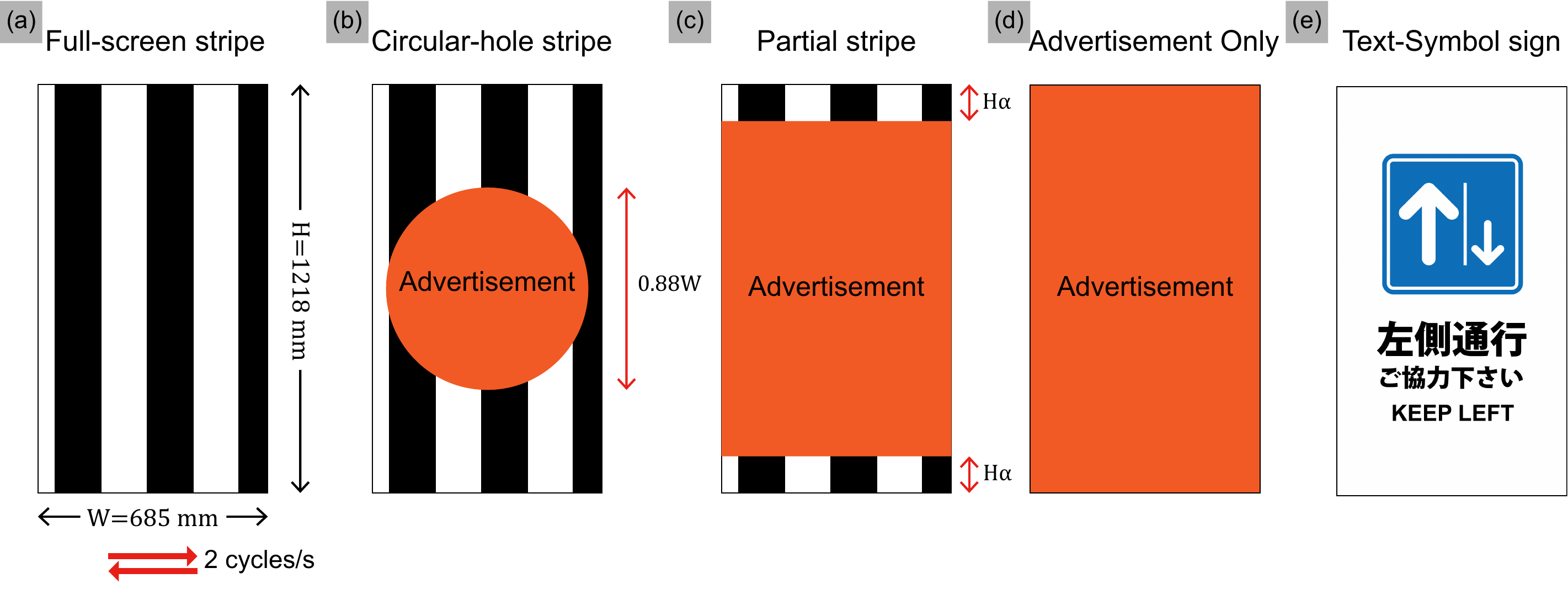}
    \caption{The five visual stimulus types presented on each MDS display ($W=685$ mm, $H=1218$ mm). (a) Full-screen stripe: a black-and-white stripe pattern covering the entire display. (b) Circular-hole stripe: a circular advertisement region (diameter $0.88W$) placed at the center of the stripe pattern. (c) Partial stripe: an advertisement region sandwiched between stripe bands at the top and bottom of the display. (d) Advertisement-only (AD): the advertisement content only, with no stripe pattern, serving as a control condition. (e) Text-symbol sign: a conventional sign combining directional text and arrow symbols instructing pedestrians to keep to one side (left-side example shown; a mirrored right-side version was also used). In (a)-(c), the stripe pattern moved laterally at 2 cycles/s.}
    \label{fig:stimuli}
\end{figure}

We primarily used simple moving stripes to vary stimulus parameters systematically while observing pedestrian behavior in a public setting. Figure~\ref{fig:stimuli} shows the five stimulus types. Each display measured approximately $W=685$ mm in width and $H=1218$ mm in height. All stripe stimuli contained three spatial cycles across the display width ($W$) and moved laterally at 2 cycles/s. At a viewing distance of 3.5 m, these parameters corresponded to approximately 0.27 cycles/deg (3.73$^\circ$/cycle), a linear speed of 456.7 mm/s on the 55-inch display, and an angular velocity of $7.43^\circ/\mathrm{s}$.

\paragraph{Full-screen stripe (Full).}
A moving stripe pattern covered the entire display (Fig.~\ref{fig:stimuli}(a)), providing the full-coverage condition against which reduced-coverage designs were examined. Four nominal contrast levels were prepared for Full: 100\%, 75\%, 50\%, and 25\%. These contrast levels were defined by the stimulus image values rather than by measured display luminance contrast. For the stripe patterns, pixel output values of 0, 0.25, 0.5, and 0.75 were gamma-corrected using a gamma value of 2.2 before presentation on the displays.

\paragraph{Circular-hole stripe (Circle).}
A circular advertisement region with diameter $0.88W = 602.8$ mm occupied the center of the stripe pattern (Fig.~\ref{fig:stimuli}(b)). This diameter subtended approximately $10^\circ$ at a viewing distance of 3.5 m. We used Circle to examine whether motion surrounding an advertisement was associated with pedestrian positions different from those under full-screen motion; when the advertisement was viewed, the stripes extended into peripheral vision.

\paragraph{Partial stripe (Partial).}
Stripe bands at the top and bottom of the display surrounded a vertically oriented advertisement (Fig.~\ref{fig:stimuli}(c)), leaving more space for advertising while reducing motion coverage. The height of each band was $H\alpha$: $\alpha = 9$\% in TC1 (109.6 mm on the 55-inch display) and $\alpha = 10$\% in TC2 (121.8 mm). At 3.5 m, these heights subtended approximately $1.79^\circ$ and $1.99^\circ$, respectively, a difference of $0.2^\circ$. We used Partial to examine pedestrian positions and subjective responses when motion was confined to these bands.

\paragraph{Advertisement only (AD).}
Only advertisement content was displayed, without stripes (Fig.~\ref{fig:stimuli}(d)). AD served as the TC2 control for assessing position differences associated with the directional conditions.

\paragraph{Text-symbol sign (Sign).}
Text and arrow symbols explicitly instructed pedestrians to keep to the left or right (Fig.~\ref{fig:stimuli}(e)). Sign provided a reference condition for the magnitude of explicit directional guidance.

\subsubsection{Stimuli used in TC1 and TC2}\label{sec:stimuli-by-test-case}

TC1 used only leftward-moving stripes: Full at four contrast levels to examine pedestrian responses to differences in stripe contrast, and Full, Circle, and Partial to examine responses to different coverage patterns at two depth spacings between signage displays. TC2 used both leftward and rightward versions of Full and Partial to explore motion direction and coverage pattern in bidirectional traffic, together with AD as the control and Sign as the explicit-guidance reference. The advertisements embedded in Circle and Partial in TC1 were static images, whereas TC2 used Miraikan promotional videos.

\subsection{Spatial and behavioral design}

\begin{figure}[t]
    \centering
    \includegraphics[width=0.85\linewidth]{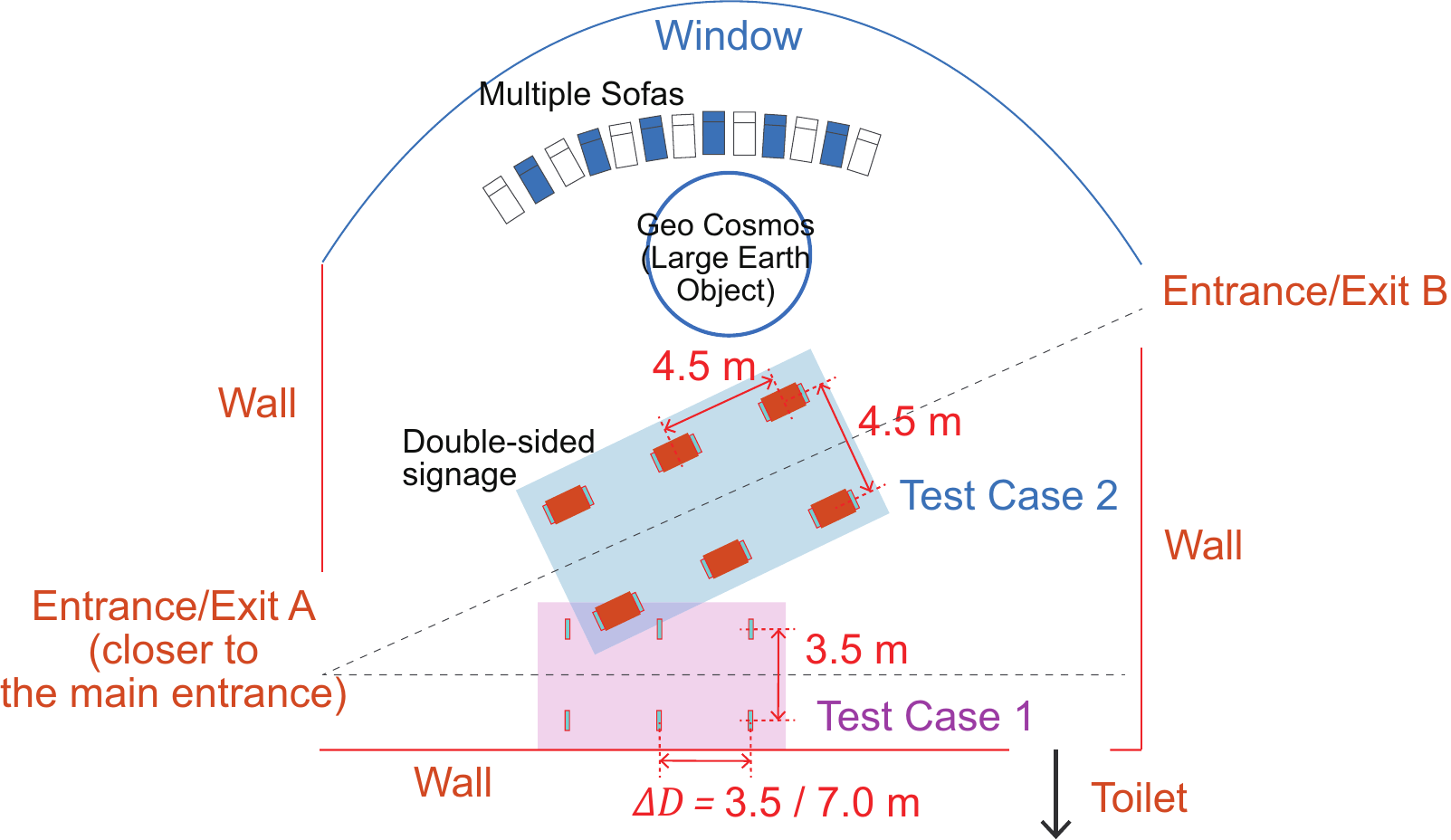}
    \caption{Floor plan of the museum's free-admission area, overlaid with the two MDS configurations. TC1 (purple) used six forward-facing displays with a lateral center-to-center spacing of 3.5 m, and a depth spacing of $\Delta D=3.5$ or $7.0$ m. TC2 (blue) used twelve displays, arranged in six back-to-back pairs with lateral and depth center-to-center spacings of 4.5 m, aligned along the straight line connecting the two Entrance/Exit points, closer to the window and to the Geo-Cosmos. The two Entrance/Exit points and the multiple sofas near the window are also indicated.}
    \label{fig:space}
\end{figure}
\begin{table}[t]
    \centering
    \begin{tabular}{lccc}
        \toprule
        \multirow{2}{*}{Characteristics} & \multirow{2}{*}{Test Case 1} & \multicolumn{2}{c}{Test Case 2}  \\ 
        & & Forward & Backward \\ \cmidrule(lr){1-1} \cmidrule(lr){2-2} \cmidrule(lr){3-4}
        Lateral display spacing & 3.5 m & \multicolumn{2}{c}{4.5 m}  \\
        Distance to the large window & Far & \multicolumn{2}{c}{Close} \\
        Ceiling & Ceiling & \multicolumn{2}{c}{Atrium} \\
        Traffic flow & One-way & \multicolumn{2}{c}{Bidirectional}\\
        Distance from Geo-Cosmos & Far & \multicolumn{2}{c}{Near} \\
        \shortstack[l]{Apparent scene brightness\\(qualitative)} & Dim & Bright & Slightly dim \\
        Spatial asymmetry & Left-side open & Left-side open & Right-side open \\
        \botrule
    \end{tabular}
    \caption{Qualitative comparison of the spatial characteristics of Test Case 1 and Test Case 2. For Test Case 2, characteristics that differed between walking directions are further broken down by the Forward and Backward directions.}
    \label{tbl:case-characteristics}
\end{table}
This study was conducted in a publicly accessible, free-admission area of the National Museum of Emerging Science and Innovation (Miraikan) in Japan. Fig.~\ref{fig:space} presents a floor plan of the area overlaid with the MDS configurations used in the two test cases. This area is one of the museum’s most iconic spaces, featuring the Geo-Cosmos, a large globe suspended from the ceiling of the atrium, as a particularly prominent visual landmark. Entrance/Exit A at the lower left is closest to the museum’s main entrance, and many visitors enter the space through this point. By contrast, Entrance/Exit B at the upper right connects the space to another area of the museum. Because this adjacent area also has its own entrance gate from outside the museum, Entrance/Exit B is used not only by visitors returning from the adjacent area but also by newly arriving visitors.

The experiment was conducted using two spatial configurations within this area. In TC1, six forward-facing signage displays were installed near the wall opposite the windows. Because the displays were arranged for viewing from only one side, the experiment was designed primarily for one-way pedestrian traffic. However, pedestrians walking in the reverse direction were not stopped as long as they did not obstruct the experiment. In TC2, 12 signage displays were installed along the line connecting the two Entrances/Exits to investigate how the MDS space affected bidirectional pedestrian traffic. The spatial configuration was designed to actively encourage movement in both directions, and the signage displays were arranged back-to-back to provide display surfaces on both sides. Photographs of the signage installations are shown in Figs.~\ref{fig:env1}(b) and \ref{fig:env2}(b, c).

Table~\ref{tbl:case-characteristics} summarizes qualitative differences between the two spaces. In the Forward direction of TC2, the windows were behind and beyond the MDS (Fig.~\ref{fig:env2}(b)); in the Backward direction, the museum wall formed its background (Fig.~\ref{fig:env2}(c)), as in TC1 (Fig.~\ref{fig:env1}(b)).

\subsection{Apparatus, LiDAR acquisition, and processing}\label{sec:lidar-acquisition}
Both test cases used 55-inch, vertically oriented displays (JAPANNEXT JN-IPS5503TUHDR, 1920×1080 px, display area: W685 mm × H1218 mm). To measure pedestrian flow, a LiDAR sensor (Hesai Technology JT128) with a hemispherical field of view was suspended from the ceiling at approximately 4.5 m above the floor in both test cases. The experiments were conducted from 10:00 to 17:00, in accordance with the museum's operating hours, with continuous image presentation on the MDS and continuous point-cloud acquisition by the LiDAR. The number and arrangement of displays and the corresponding measurement areas differed between the cases, as described below.

Both test cases used a common core LiDAR processing pipeline, with processing details and outcome definitions provided in the Supplementary Information (SI; Sec.~\ref*{si-sec:point-cloud-processing}; Fig.~\ref*{si-fig:proc_pcl}). Point clouds were transformed to corridor-aligned coordinates with the origin at the center of the MDS space, cropped to the measurement region, and filtered to remove fixed artifacts. Pedestrian positions were approximated by cluster centroids. Each centroid was classified as Forward, Backward, Stationary ($XY$ displacement below 0.05~m per frame), Ambiguous (neither directional criterion met), or No direction (no preceding-frame correspondence) according to defined criteria based on adjacent-frame correspondences. Using these directional labels and relative motion, we detected passing events between oppositely moving pedestrians. For the position analyses, observations within a window extending 100 frames before and after each event's representative crossing frame were excluded. To summarize pedestrian spatial distributions during one-way flow, frame-wise median $X$ and $Y$ positions were computed for each analyzed walking direction and averaged within fixed, nonoverlapping 100-frame windows (10 s at 10 fps) anchored to the original frame indices within each day. Each average used only frames with valid observations, without interpolating missing frames; windows with fewer than 10 such frames were excluded. Each retained window represented an aggregate position rather than an individual trajectory. Processing details and the rationale for the 10-s interval are provided in the SI.

\subsection{Participants, ethics, and procedure}\label{sec:general-procedure}
All pedestrians who passed through the MDS space during the experimental period were included as participants. Staff manually counted pedestrians, and age and gender were self-reported by questionnaire respondents. In keeping with the study's aim of targeting naturally occurring pedestrian flow in a real-world space, no eligibility restrictions were imposed based on walking ability or vision. Because pedestrian movement was recorded solely using LiDAR point-cloud data, participants’ ages and genders were not individually identified.

Individual written consent forms were not collected because the experiment was conducted as a large-scale field trial in a public, free-admission area of a national museum. Instead, a notice-based opt-out consent procedure approved by the Ethics Committee of NTT Communication Science Laboratories (protocol R07-013) was used. The study was conducted in accordance with the ethical standards of the Declaration of Helsinki. An information notice was displayed before the MDS space, informing visitors about the experiment, the data to be collected, and the handling of personal information. Visitors who did not wish to participate could use the adjacent avoidance route. By passing beyond the information notice and entering the MDS space, visitors were regarded as having agreed to participate. The notice also stated that visitors could inform the on-site staff in order to opt out. In such cases, staff members would guide them to an adjacent avoidance route outside the measurement area or arrange for deletion or exclusion of data recorded during the relevant time window. (No participant made such a request in either TC1 or TC2.)

The experimental procedure was as follows: Staff informed visitors about the ongoing field experiment and invited them to walk through the space. No restrictions or explicit instructions were imposed regarding starting position or destination; participants simply walked toward their own intended destinations, thereby preserving naturalistic pedestrian flow. Case-specific staff arrangements and the encouragement of bidirectional walking in TC2 are described in the respective test-case sections.

After passing through the area, participants could voluntarily complete a questionnaire using on-site iPads. The questionnaire was created using Typeform \citep{typeform} and was available only on these iPads; no public link was distributed. An explanation booth provided visitors with information about the research after passage. The iPad arrangements and booth positions are described separately for each case.

\subsection{Questionnaire}\label{sec:questionnaire}

The voluntary post-experience questionnaire was intended to complement the objective LiDAR-based behavioral measurements described above by assessing visitors' subjective experiences, including awareness of the visual stimuli, perceived influence on walking, and discomfort during exposure. It comprised ten items developed for this field study: five focal items and five non-focal items included to make the questionnaire's focus less apparent (SI, Sec.~\ref*{si-app:questionnaire-items}, Table~\ref*{si-tbl:questionnaire-items}). These ten items were followed by demographic questions on gender and age group. The focal items addressed Motion awareness, Walking against motion, Walking influence, Following floor lines, and content visibility. Content visibility was assessed using a four-category adaptation of the Perceptual Awareness Scale (PAS) \citep{ramsoy2004introspection}. The non-focal items addressed fatigue, sound awareness, awareness of other pedestrians, audiovisual harmony, and sickness while walking. Among these ten items, all except PAS used a five-point Likert scale (1--5); PAS used four categories (1--4).

Focal and non-focal items were intermixed to reduce responses based on inferred expectations \citep{orne1962social,nichols2008good,iarygina2025demand}. Although Sickness was a non-focal item, it provides practically relevant information about discomfort associated with the stimulus designs. Items were assessed separately rather than as a common psychological construct.

\section{Test Case 1: predominantly one-way pedestrian flow}

TC1 addressed the following research question in predominantly one-way museum visitor flow:

\begin{itemize}
    \item \textbf{RQ1.} \textit{Do pedestrian lateral-position distributions differ across stripe contrast levels and coverage patterns, and are coverage-related trends similar across two display depth spacings?}
\end{itemize}
To answer RQ1, TC1 comprised two sub-experiments: one compared four Contrast levels using Full-screen stripes, and the other compared three coverage patterns (CovPat: Full, Circle, and Partial) across two depth spacings.

\subsection{Environment}\label{sec:exp1-environment}

TC1 was conducted in a relatively dim area of the free-admission space farther from the large windows.

\begin{figure}
    \centering
    \includegraphics[width=\linewidth]{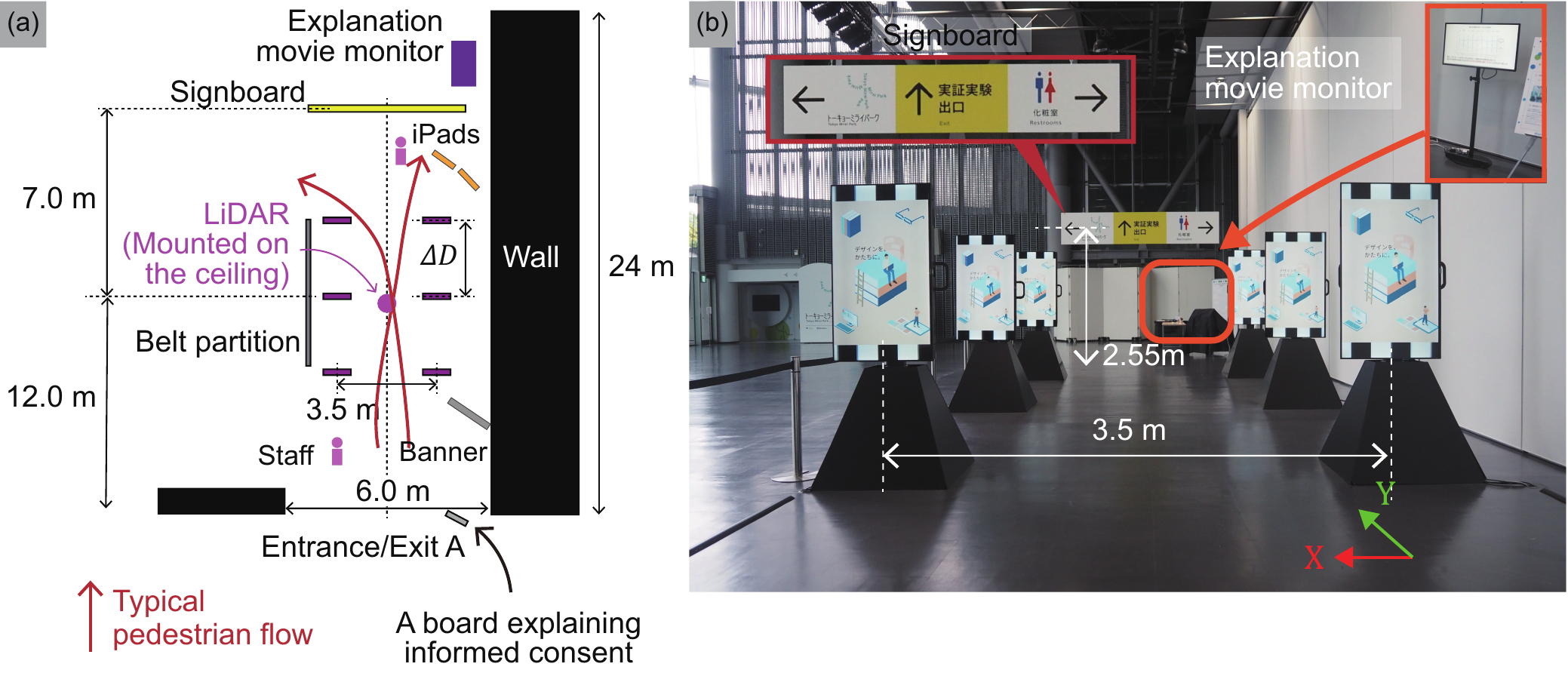}
    \caption{Experimental environment for Test Case 1. (a) Schematic layout showing the six signage displays, the LiDAR sensor and its field of view, the belt partition, the signboard, the explanation movie monitor, the informed-consent notice board, staff positions, and the entrance. (b) Photograph of the installed signage displays viewed from the entrance, showing the explanation movie monitor and the $XY$ coordinate system used for the pedestrian positional analysis.}
    \label{fig:env1}
\end{figure}

The layout shown in Fig.~\ref{fig:env1}(a) was identical to that used in our previous study \citep{mikawa-vection2026}.
Six displays of the model described in Sec.~\ref{sec:lidar-acquisition} were arranged with a lateral center-to-center spacing of 3.5 m. In the depth direction, two depth spacing conditions were used: $\Delta D=3.5$ m (Small) and $\Delta D=7.0$ m (Large). Hereafter, we refer to the area defined by these six displays (3.5 m laterally and $2\Delta D$ in depth) as the MDS space.
The MDS was open on the left side and bounded on the right by a wall located 1.75 m lateral to the rightmost display. This layout followed the existing spatial design and visitor circulation plan. A custom-made signboard (3.5 m wide, 0.7 m high) was suspended with its center 2.55 m above the floor. The signboard, resembling a station sign, indicated three possible destinations within the space: other exhibitions to the left, the explanation movie monitor straight ahead, and the restrooms to the right.
The distance from the center of the MDS space to the entrance was 12.0 m, and the distance to the signboard was 7.0 m (with the Large depth spacing, the rear displays were located directly beneath the signboard). 

An explanation booth with a movie monitor was installed along the rear wall to provide visitors with information about the research after they passed through the MDS space (Fig.~\ref{fig:env1}(a)). The booth was visible from the entrance (Fig.~\ref{fig:env1}(b)), introducing asymmetry in the spatial layout and visual field, but its position and the spatial arrangement were unchanged across days.
The experiment was conducted in a space with large windows, and sound from events in the adjacent area was sometimes present; however, these conditions were similar across measurement days.

Two staff members were stationed at the front and rear of the experimental area. Three iPads were prepared for participants to answer the questionnaire: two (Apple MD4A4J/A, 11”) were mounted on stands in the area beyond the passage route, and a rear staff member carried the third iPad (Apple MPQ03J/A, 10.9”). These iPads provided the voluntary post-passage questionnaire described in Sec.~\ref{sec:general-procedure}.

\subsection{Stimuli and conditions}\label{sec:exp1-stimuli}

TC1 comprised two sub-experiments. The Contrast experiment compared four levels (100\%, 75\%, 50\%, and 25\%) using Full-screen stripes at Small depth spacing. The coverage-pattern and depth-spacing experiment compared Full, Circle, and Partial at both Small ($\Delta D = 3.5$ m) and Large ($\Delta D = 7.0$ m) depth spacings, using black-and-white stripes at 100\% contrast. Full occupied the entire display, Circle embedded an advertisement image within a central circular opening, and Partial placed stripes along the top and bottom display edges. Thus, the first sub-experiment explored contrast differences, whereas the second explored coverage-pattern differences within each depth spacing.

Spatial frequency and stripe speed were as described in Sec.~\ref{sec:visual-stimuli-design}. All stripes moved leftward continuously throughout the museum's seven-hour opening period. TC1 did not include a static-stripe baseline.

A single condition was presented throughout each calendar day to avoid visitors encountering display transitions. The Full condition at Small depth spacing was shared by both sub-experiments. The experiment spanned nine days: six weekdays and three weekend or holiday days. All four Contrast conditions were presented on weekdays. For the coverage-pattern and depth-spacing experiment, Full-Small, Circle-Large, and Partial-Large were presented on weekdays, and Full-Large, Circle-Small, and Partial-Small on weekends/holidays.
The layout and scheduled events remained essentially unchanged across days, but pedestrian volume varied (Fig.~\ref{fig:user-num1}(a)).

\subsection{Participants and procedure}\label{sec:exp1-participants}

\begin{figure}
    \centering
    \includegraphics[width=\linewidth]{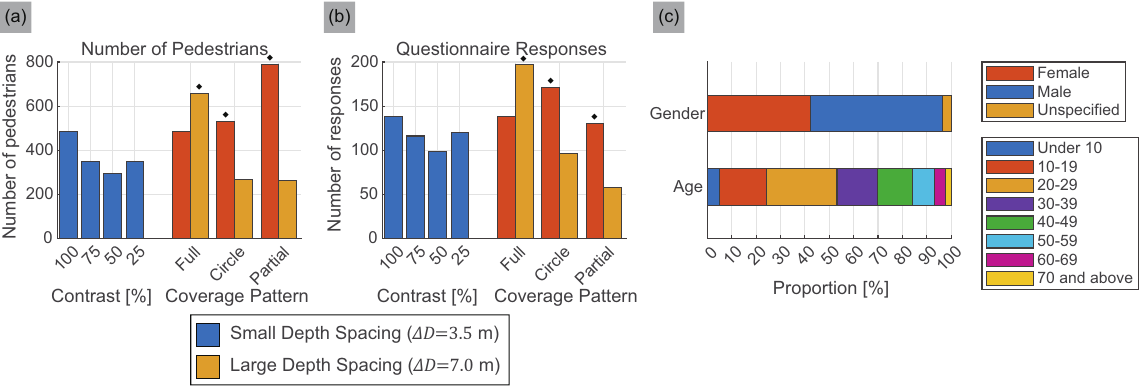}
    \caption{Participant numbers and demographics for Test Case 1. (a) Number of pedestrians passing through the MDS space, manually counted by staff, for each contrast condition (25–100\%) and each CovPat condition (Full, Circle, Partial), split by small ($\Delta D=3.5$ m) and large depth spacing ($\Delta D=7.0$ m). (b) Number of questionnaire responses collected via iPads for the same conditions. (a,b) Note that the Contrast-100\% and Full-Small conditions refer to the same experimental day. Conditions conducted on weekends/holidays are marked with $\blacklozenge$. (c) Gender and age distribution of the questionnaire respondents.}
    \label{fig:user-num1}
\end{figure}

TC1 involved 3,982 pedestrians over nine experimental days. Participant inclusion followed Sec.~\ref{sec:general-procedure}. Figure~\ref{fig:user-num1}(a) shows the daily pedestrian counts recorded manually by staff, and Fig.~\ref{fig:user-num1}(b) shows the daily numbers of questionnaire responses collected via iPads. The mean daily pedestrian count was 442.4$\pm$186.3, and the mean daily number of questionnaire responses was 124.9$\pm$42.3 (mean $\pm$ SD).

The demographics (age and gender) of the questionnaire respondents are shown in Fig.~\ref{fig:user-num1}(c).
The age distribution of respondents was as follows: under 10, 5.06\%; 10--19, 19.36\%; 20--29, 28.76\%; 30--39, 16.70\%; 40--49, 14.30\%; 50--59, 8.87\%; 60--69, 4.70\%; and 70 and over, 2.24\%. The gender distribution was 42.28\% female, 54.28\% male, and 3.44\% unspecified. Participants in their 20s accounted for the largest proportion of the sample.

TC1 followed the common procedure in Sec.~\ref{sec:general-procedure}. An agency staff member stood near the entrance to inform visitors about the ongoing field experiment and to invite them to walk through the space. If a participant stopped midway, staff standing on the left side of the MDS space encouraged them to continue walking. Because the space functioned as an exhibition area, visitors were not prevented from walking in the opposite direction; however, such individuals were excluded from the pedestrian point-cloud data (see Sec.~\ref{sec:exp1-analysis}). No agency staff members entered the MDS space during the experiment. The information notice for TC1 was displayed before the MDS space (bottom-right of Fig.~\ref{fig:env1}(a)), following the common consent procedure in Sec.~\ref{sec:general-procedure}.

\subsection{Pedestrian analysis}\label{sec:exp1-analysis}
The LiDAR data were processed using the common pipeline outlined in Sec.~\ref{sec:lidar-acquisition} and detailed in the SI (Sec.~\ref*{si-sec:point-cloud-processing}), with the following TC1-specific settings.

The nominal measurement region was defined laterally by the inner edges of the displays on either side; given their 3.5-m center-to-center spacing and 0.685-m display width, this yielded an interior width of approximately 2.8 m. In the depth direction, the layout-based region extended from 1.5 m in front of the frontmost displays to 1.5 m behind the rearmost displays, with a vertical extent of 4.5 m. With Small depth spacing between displays ($\Delta D=3.5$ m), these nominal dimensions were 10.0 m $\times$ 2.8 m $\times$ 4.5 m (depth $\times$ width $\times$ height).
Under Large depth spacing ($\Delta D=7.0$ m), the corresponding layout-based depth was 17 m (2 $\times$ 7.0 m + 2 $\times$ 1.5 m). The nominal depth ranges were $-5\leq Y\leq5$ m for Small spacing and $-8.5\leq Y\leq8.5$ m for Large spacing. Static artifacts were removed from both layouts during preprocessing (SI, Sec.~\ref*{si-sec:preprocessing}). We analyzed coverage patterns separately for the two depth spacings to accommodate their different spatial extents. Only centroids classified as Forward were retained and processed using the common passing-event exclusion and 10-s aggregation procedures (Sec.~\ref{sec:lidar-acquisition}).

For the primary analyses, we used Type III analysis of variance (ANOVA). The Contrast model included Contrast Level alone ($X \sim \mathrm{ContrastLevel}$). For coverage pattern, we fitted a separate one-factor model, $X \sim \mathrm{CovPat}$, for each depth spacing, comparing Full, Circle, and Partial within that layout. Holm-adjusted pairwise comparisons followed the omnibus tests, with the three CovPat comparisons adjusted as one family within each spacing. %

To describe the weekday and weekend/holiday measurements, we summarized lateral position by measurement day and day type (SI, Table~\ref*{si-tab:tc1-day-descriptives}). Supplementary analyses divided each depth range into three equal intervals (Front, Middle, and Rear), as detailed in SI Secs.~\ref*{si-app:tc1-contrast-depth-zones} and \ref*{si-app:tc1-spacing-specific}. %

\subsection{Results of the pedestrian analysis}

\begin{figure}
    \centering
    \includegraphics[width=\linewidth]{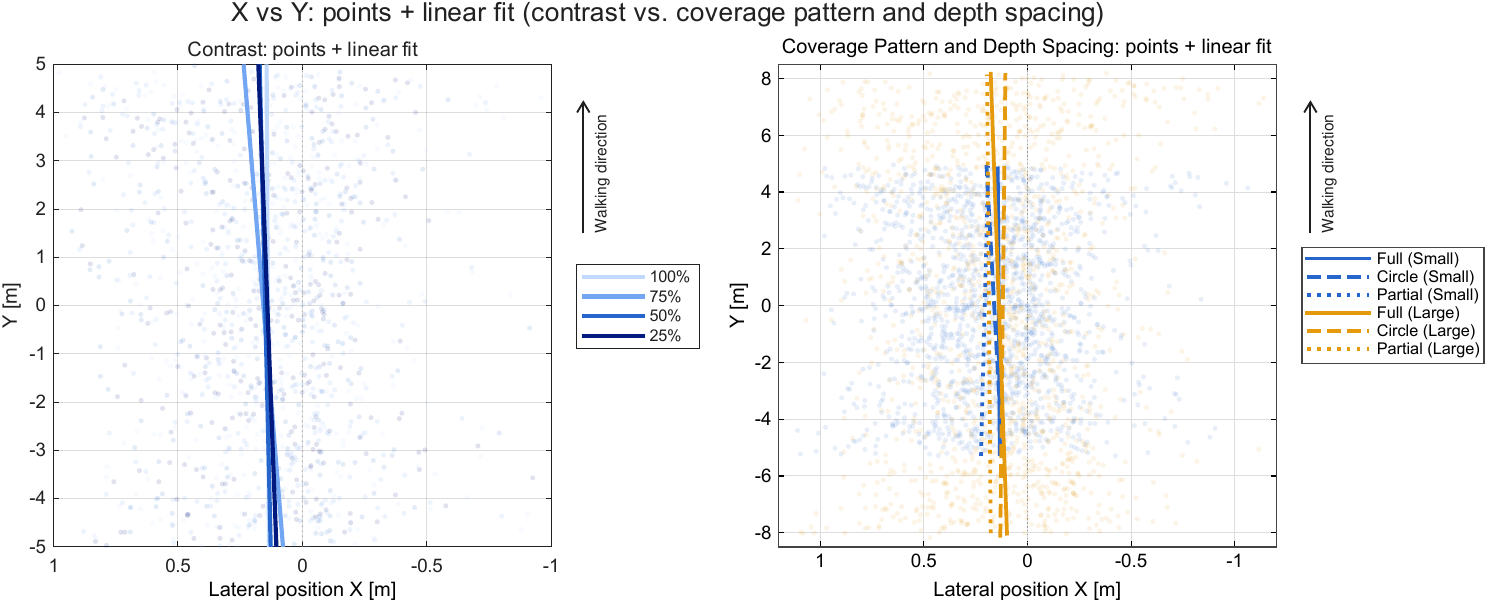}
    \caption{Lateral ($X$) position versus depth ($Y$) in TC1. Each point is the mean of frame-wise median centroid positions within a retained 10-s window containing data in at least 10 frames. (Left) The four Contrast conditions. (Right) Three coverage pattern (CovPat) conditions at both Small and Large depth spacings, using the same retained blocks as the spacing-specific analyses. The nominal depth ranges were $-5\leq Y\leq5$ m for Small spacing and $-8.5\leq Y\leq8.5$ m for Large spacing (SI, Sec.~\ref*{si-sec:preprocessing}). Lines are descriptive linear fits of $X$ on $Y$ using all retained blocks for each condition; right-panel lines are restricted to each condition's observed depth range. Positive $X$ is to the left when facing Forward, in the direction of stripe motion. The plotted points summarize spatial distributions rather than individual displacement.}
    \label{fig:trajectory1}
\end{figure} %

\begin{figure}
    \centering
    \includegraphics[width=\linewidth]{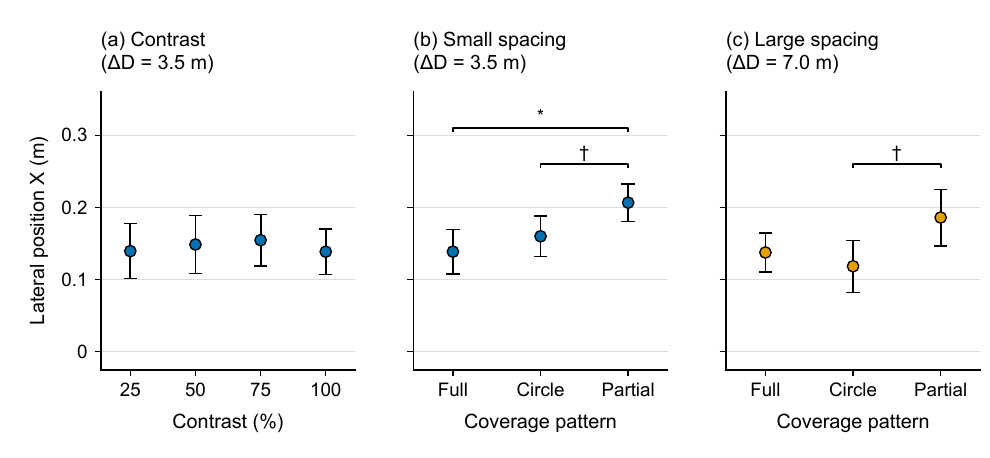}
    \caption{Estimated marginal means of lateral ($X$) position in TC1, with pointwise 95\% confidence intervals: (a) Contrast, (b) CovPat at Small spacing, and (c) CovPat at Large spacing over the 17-m range. Positive $X$ is leftward when facing Forward, in the stripe-motion direction. (b,c) Brackets indicate Holm-adjusted comparisons within each spacing ($^\dagger p<.05$, $^*p<.01$; three CovPat comparisons per family).}
    \label{fig:res1}
\end{figure}

We examined whether pedestrians' lateral ($X$) positions differed across experimental conditions using the 10-s aggregate positions (defined in SI Sec.~\ref*{si-sec:point-cloud-processing}). The primary analyses compare lateral-position distributions over the full sampled depth range. Each observation summarizes a spatial distribution within a time window rather than the trajectory of an individual pedestrian.

Descriptively, Forward centroid observations outnumbered Backward observations at both depth spacings. Counts and classifications for the nominal 10-m Contrast and Small-spacing data and the 17-m Large-spacing data are reported in SI Sec.~\ref*{si-app:pedestrian-data-summary}.

\subsubsection{Contrast condition}\label{sec:exp1-contrast}
The position analysis used 1,314 valid 10-s blocks. Figures~\ref{fig:trajectory1} (left) and \ref{fig:res1}(a) summarize lateral positions across depth and the condition means. No clear condition differences were detected among the four Contrast levels. The one-way ANOVA revealed no significant effect of Contrast Level ($F(3,1310)=0.182$, $p=.909$, $\eta^2_p<.001$). Holm-adjusted pairwise comparisons among the four contrast levels showed no significant differences (all $p_{\mathrm{Holm}}=1$). The supplementary depth-zone analyses likewise detected no significant effect of Contrast Level in Front, Middle, or Rear, consistent with the full-range result (SI, Sec.~\ref*{si-app:tc1-contrast-depth-zones}).

\subsubsection{Coverage pattern conditions}\label{sec:exp1-cov-depth}
In Small depth spacing, the position analysis used 1,558 valid 10-s blocks (Full: 427; Circle: 518; Partial: 613; Fig.~\ref{fig:res1}(b)). The mean lateral positions followed the order Partial, Circle, and Full, from most to least positive (Figs.~\ref{fig:trajectory1} and \ref{fig:res1}(b)).
The one-way ANOVA showed a significant CovPat effect ($F(2,1555)=6.092$, $p=.002$, $\eta^2_p=.008$). Holm-adjusted comparisons showed that Partial was more positive than Full (estimated difference $=6.8$ cm, $t(1555)=3.321$, $p_{\mathrm{Holm}}=.003$, Cohen's $d=.209$) and Circle ($+4.7$ cm, $t(1555)=2.408$, $p_{\mathrm{Holm}}=.032$, $d=.144$). The Circle--Full difference was not significant ($+2.1$ cm, $t(1555)=1.004$, $p_{\mathrm{Holm}}=.315$, $d=.066$). Because Full was measured on a different day type from Circle and Partial, comparisons involving Full should be interpreted cautiously. In the supplementary depth-zone analyses, Partial was significantly more positive than Full and Circle in Front, and more positive than Circle in Middle; no pairwise comparison was significant in Rear (SI, Sec.~\ref*{si-app:tc1-small-coverage}).

In Large depth spacing, the 17-m position analysis used 1,422 valid 10-s blocks (Full: 703; Circle: 389; Partial: 330; Fig.~\ref{fig:res1}(c)). CovPat was significant ($F(2,1419)=3.270$, $p=.038$, $\eta^2_p=.005$). Partial was more positive than Circle (estimated difference $=6.8$ cm, $t(1419)=2.482$, $p_{\mathrm{Holm}}=.040$, Cohen's $d=.186$), whereas Partial--Full ($+4.8$ cm, $p_{\mathrm{Holm}}=.092$, $d=.133$) and Full--Circle ($+1.9$ cm, $p_{\mathrm{Holm}}=.406$, $d=.053$) were not significant. Descriptively, the fitted line for Full became more positive with depth, exceeding Circle and approaching Partial toward the rear (Fig.~\ref{fig:trajectory1}, right). Because Full was measured on a different day type from Circle and Partial, comparisons involving Full should be interpreted cautiously.
In the supplementary depth-zone analyses, the only significant pairwise contrast within the three equal depth intervals was Partial--Full in the middle third (SI, Sec.~\ref*{si-app:tc1-large-coverage}). %

Overall, smaller stripe coverage was associated with more positive lateral positions at Small spacing. At Large spacing, the pattern was less uniform: the descriptive Full--Partial gap narrowed toward the rear. This suggests that the relative usefulness of smaller and larger stripe regions may depend on the extent of the walking space.
\subsection{Questionnaire response analysis}

The questionnaire response plots are shown in SI Fig.~\ref*{si-fig:response1}, and the analysis methods and detailed results are provided in SI Sec.~\ref*{si-sec:questionnaire-tc1}. Most questionnaire items showed no significant differences between conditions, but among the focal items, visibility ratings were higher at 100\% than at 25\% Contrast, and Motion awareness ratings were lower with Circle and Partial than with Full. Sickness, a non-focal item, was also rated lower with Circle and Partial than with Full.

\subsection{Summary for Test Case 1}

TC1 detected no significant differences in pedestrian lateral position among the Contrast levels. At Small depth spacing, Partial showed a significantly greater positional bias in the stripe-motion direction than Full and Circle. At Large depth spacing, Partial differed significantly from Circle but not Full. Smaller stripe coverage thus tended to be associated with a greater bias, although effect sizes were small and the descriptive Full--Partial gap narrowed toward the rear of the longer layout. Full received higher Motion awareness and Sickness ratings than Partial. These findings motivate examining designs that confine moving stripes to the display edges while retaining a central area for advertisements. %

\section{Test Case 2: bidirectional pedestrian flow}
Building on TC1, we conducted a similar experiment in TC2 in the same free-admission area, with the MDS positioned closer to the windows and arranged to permit bidirectional walking. The practical aim was to examine whether visual motion on signage could encourage pedestrians to keep left or right, including when passing one another. We addressed two research questions:

\begin{itemize}
    \item \textbf{RQ2.} \textit{Do pedestrian lateral-position distributions differ with stripe-motion direction (L/R) and coverage pattern (Full/Partial)?}
    \item \textbf{RQ3.} \textit{Does passing-side choice differ between the stripe conditions and the advertisement-only control (AD)?}
\end{itemize}

In practical terms, RQ2 was addressed by comparing lateral-position distributions among the stripe conditions and with the advertisement-only control (AD) during one-way flow in each walking direction (Secs.~\ref{sec:exp2-oneway-3way} and \ref{sec:exp2-oneway-dunnett}). For RQ3, we tested whether the probability of left-side passing differed between the stripe conditions and AD (Sec.~\ref{sec:exp2-passing}). Lateral display spacing was widened from 3.5 m in TC1 to 4.5 m in TC2, and depth spacing was also fixed at 4.5 m. Actual bidirectional encounters were relatively sparse: only 1.84--9.32\% of frames containing centroid observations were classified as bidirectional flow. We therefore used the extracted one-way-flow data to examine lateral-position modulation, as in TC1, while also testing TC2's original question of whether stripe motion could guide passing-side choice during encounters.

\subsection{Environment}

\begin{figure}[t]
    \centering
    \includegraphics[width=\linewidth]{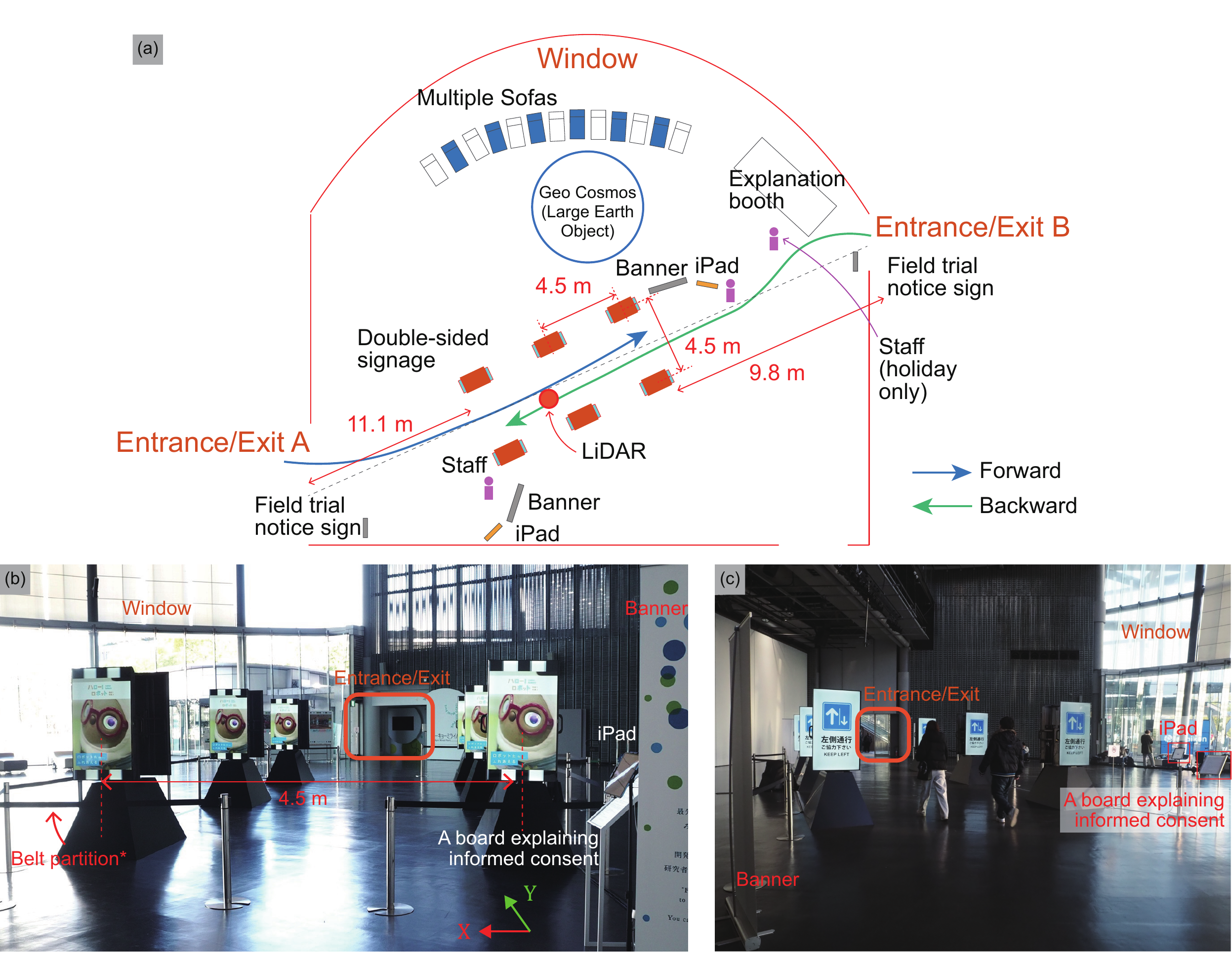}
    \caption{Experimental environment for Test Case 2. (a) Schematic layout showing the twelve signage displays, arranged as six back-to-back pairs, connecting the two Entrance/Exit points. (b) Photograph of the signage displays viewed from Entrance/Exit A (closer to the museum's main entrance; Forward direction), with the $XY$ coordinate system used for the pedestrian positional analysis. The photographs were taken while the museum was closed, with a belt partition in front of the signage; during the experiment, it was moved aside. (c) Photograph of the signage displays viewed from Entrance/Exit B (Backward direction).}
    \label{fig:env2}
\end{figure}

The experimental setup is illustrated in Fig.~\ref{fig:env2}(a). Although this test case was conducted in the same space as TC1 and used the same entrance (Entrance/Exit A) as the starting point of the space, the twelve displays were arranged to connect this entrance with another entrance located on the opposite side of the space (Entrance/Exit B), closer to the windows. Twelve signage displays of the model described in Sec.~\ref{sec:lidar-acquisition} were prepared. They were grouped into six back-to-back pairs and installed with a center-to-center distance of 4.5 m between adjacent pairs. Consequently, the MDS configuration was located closer to the windows than in TC1, and the MDS space itself remained brighter throughout the entire experimental period. In addition, because the layout was designed to allow bidirectional pedestrian traffic, the lateral display spacing was wider than that in TC1.

The lower-left entrance (Entrance/Exit A) in Fig.~\ref{fig:env2}(a) was much closer to the main entrance of the museum, and pedestrian traffic entering from it was thus substantially greater than that entering from the other entrance. Thus, staff members invited pedestrians to enter from the upper-right entrance (Entrance/Exit B).

At each entrance located at either end of the MDS space, an iPad, a banner, and signs providing information about the field experiment and the relevant ethical considerations were installed.
The area closer to the windows also featured an atrium, where a large globe—an iconic symbol of the museum—was suspended from the ceiling and frequently attracted visitors’ attention.

\subsection{Stimuli and conditions}

Full-screen stripe and Partial stripe conditions were used to explore the roles of motion direction and coverage pattern. In Partial, the central advertisement used the museum's videos instead of the static images used in TC1, and each stripe band was slightly taller ($\alpha=10$\% rather than $\alpha=9$\%). In both stripe conditions, the pattern moved at a constant speed either rightward (Right; R) or leftward (Left; L) throughout each experimental day.

The Text-symbol sign (Sign; Fig.~\ref{fig:stimuli}(e)) instructed pedestrians to keep right or left and served as an explicit-guidance reference condition. Advertisement only (AD; Fig.~\ref{fig:stimuli}(d)), without stripes, served as the control. The experiment thus comprised $2 \times 3 + 1 = 7$ conditions: two directions for Full, Partial, and Sign, plus AD.

The experiment was conducted in January and February 2026 over a total of 16 days, seven of which were holidays. One holiday was therefore assigned to each condition. Of the remaining nine weekdays, two days were assigned to each of the two directional full-screen-stripe conditions, whereas one day was assigned to each of the other five conditions. Thus, Full-L and Full-R were each presented on three days (one holiday and two weekdays), whereas AD, Partial-L, Partial-R, Sign-L, and Sign-R were each presented on two days (one holiday and one weekday). Within these constraints, the order of conditions across the 16 days was randomly generated.

Left (L) and Right (R) in the stimulus condition labels are defined from each pedestrian's viewpoint, for both Forward and Backward walking. The corridor coordinates shown in Fig.~\ref{fig:trajectory2} retain a fixed physical frame: positive $X$ is left when facing Forward ($+Y$) and right when facing Backward ($-Y$). The inferential position analyses instead used walker-referenced lateral position $X_{\mathrm{ego}}$, positive to each pedestrian's own left (Sec.~\ref{sec:exp2-analysis}).

\subsection{Participants, procedure, and questionnaire}\label{sec:exp2-participants}
TC2 involved 14,420 pedestrians over sixteen experimental days. Participant inclusion, consent, and the experimental procedure followed Sec.~\ref{sec:general-procedure}. In TC2, bidirectional pedestrian traffic was actively encouraged, and the post-passage information booth was located farther from the MDS space than in TC1.

\begin{figure}
    \centering
    \includegraphics[width=\linewidth]{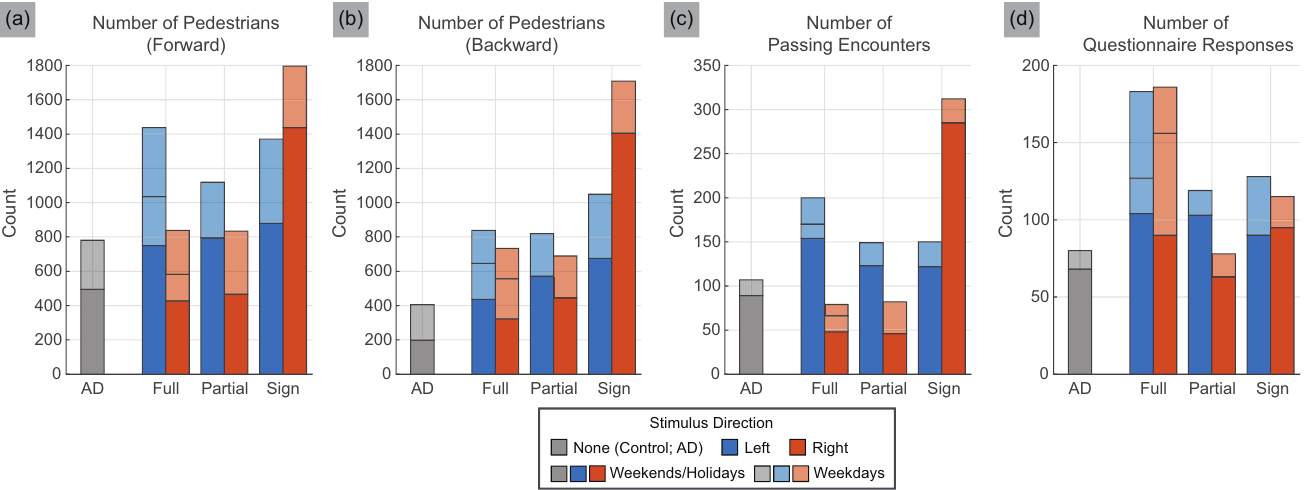}
    \caption{Participant numbers in Test Case 2 for each visual condition (AD, Full, Partial, Sign), split by Stimulus Direction (Left/Right) and day type (weekday vs. weekend/holiday). Each segment of a stacked bar represents a single day's count; because Full-L and Full-R were each conducted over three days, their bars are stacked from three daily counts, whereas the remaining conditions, conducted over two days each, are stacked from two.
    (a,b) Number of manually counted pedestrians walking in the forward (a) and backward (b) directions, respectively. (c) Number of detected passing encounters. (d) Number of questionnaire responses collected via iPads. }
    \label{fig:user-num2}
\end{figure}

\begin{figure}
    \centering
    \includegraphics[width=0.6\linewidth]{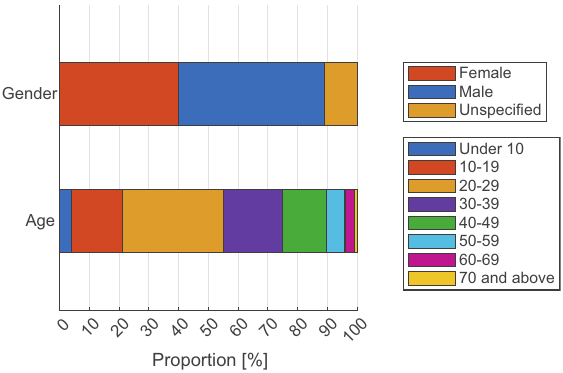}
    \caption{Demographic characteristics of the questionnaire respondents in Test Case 2: distribution of self-reported gender (Female, Male, Unspecified) and age group (Under 10 to 70 and above), expressed as a proportion of all respondents.}
    \label{fig:demographics-2}
\end{figure}

Participant counts are summarized in Fig.~\ref{fig:user-num2}. The demographic characteristics of the questionnaire respondents are shown in Fig.~\ref{fig:demographics-2}. The age distribution was: under 10, 4.16\%; 10--19, 16.87\%; 20--29, 34.20\%; 30--39, 19.57\%; 40--49, 14.96\%; 50--59, 6.19\%; 60--69, 3.26\%; and 70 and over, 0.79\%. The gender distribution was 40.04\% female, 49.16\% male, and 10.80\% unspecified.

TC2 used the questionnaire described in Sec.~\ref{sec:questionnaire} and SI Table~\ref*{si-tbl:questionnaire-items}, with the Following floor lines item revised from ``I walked along the central line on the floor'' to ``I walked along the lines on the floor.'' In TC1, floor-tile grooves formed a prominent centerline through the MDS space, whereas the TC2 layout was diagonal to the floor lines.

\subsection{Pedestrian analysis}\label{sec:exp2-analysis}
The LiDAR processing followed the common pipeline outlined in Sec.~\ref{sec:lidar-acquisition}. The cropped interior volume was approximately 3.8 m wide, reflecting the 4.5-m lateral display spacing, and 12 m deep, including the 1.5-m margins before and after the MDS space. The input coordinates used the corridor-aligned frame defined in the SI (Sec.~\ref*{si-sec:point-cloud-processing}).

The position analysis examined whether lateral distributions differed with stripe-motion direction (Left/Right) and coverage pattern (Full/Partial). The factorial model focused on these four stripe conditions, while comparison with AD tested whether they differed from a condition without lateral stripe motion. For passing-side choice, we first compared left-side passing between the stripe conditions and AD to address RQ3, then examined stimulus-congruent passing in a supplementary analysis of the four stripe conditions.

For the position analysis during one-way flow, centroids classified as Forward or Backward were retained and processed separately for each walking direction using the common passing-event exclusion and 10-s aggregation procedures (Sec.~\ref{sec:lidar-acquisition}). To express lateral position in terms of each pedestrian's own left and right, we reversed the sign of $X$ for observations classified as Backward walking before fitting the position models ($X_{\mathrm{ego}}=X$ for Forward and $X_{\mathrm{ego}}=-X$ for Backward). Positive values therefore denote each pedestrian's own left, and negative values their own right. We used linear mixed models (LMMs) for the statistical analysis, including Day (measurement session) as a random intercept and DayType (weekday vs. weekend/holiday) as a fixed effect to account for shared day-level variation in $X_{\mathrm{ego}}$. The factorial and AD-comparison position models used Type III tests with Satterthwaite degrees of freedom and estimated marginal means with Kenward--Roger degrees of freedom. Partial eta-squared ($\eta^2_p$) values were approximated from the $F$ statistics and their Satterthwaite degrees of freedom. %

As supplementary position analyses, we repeated the factorial and AD comparisons in three equal 4-m depth zones (Front, Middle, and Rear) spanning the 12-m range and defined relative to walking direction. The same LMMs were fitted to 10-s positions recomputed within each zone and walking direction (SI, Sec.~\ref*{si-sec:tc2-depth-zones}).

Passing encounters were analyzed using events as the analysis unit. The five-condition comparison with AD used $\text{Success}_{\mathrm{left}}$ to indicate left-side passing in every condition, because AD has no stripe-motion direction (Sec.~\ref{sec:exp2-passing}). A supplementary analysis of the four stripe conditions examined whether the probability of passing on the side of stripe motion differed by coverage pattern (Full/Partial) and motion direction (Left/Right) (SI, Sec.~\ref*{si-sec:passing-factorial}).

\subsection{Results of the pedestrian analysis}\label{sec:exp2-pedestrian-results}

\begin{figure}[t]
    \centering
    \includegraphics[width=\linewidth]{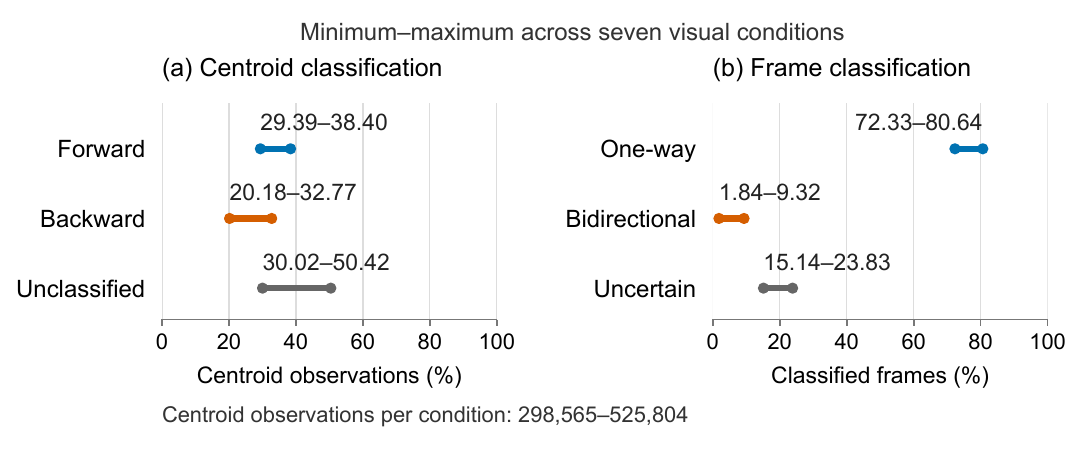}
    \caption{TC2 centroid and frame classifications across the seven recorded conditions, including Sign. Segments show the minimum--maximum percentages across conditions, with the ranges labeled numerically. (a) Centroid observations before direction-based exclusion (298,565--525,804 per condition); Unclassified comprises Stationary, Ambiguous, and No direction. (b) Point-cloud frames. Percentages use all centroids or classified frames within each condition as denominators. These are descriptive ranges, not confidence intervals; frame classification is independent of passing-event detection.}
    \label{fig:tc2-data-summary}\label{tab:tc2-data-summary}
\end{figure}

Fig.~\ref{fig:tc2-data-summary} summarizes the centroid and frame classifications in TC2. Forward centroids remained more frequent, but the two walking directions were more balanced than in TC1. The positional summaries comprised 10,656 retained 10-s blocks across all seven recorded conditions; the inferential models used the subsets described below. We report lateral positions during one-way flow and passing-side choice during encounters separately. Detailed counts and classifications are provided in the SI (Sec.~\ref*{si-app:pedestrian-data-summary}; Table~\ref*{si-tab:pedestrian-centroid-block}; Fig.~\ref*{si-fig:pedestrian-classification}(c)).

\begin{figure}[t]
    \centering
    \includegraphics[width=\linewidth]{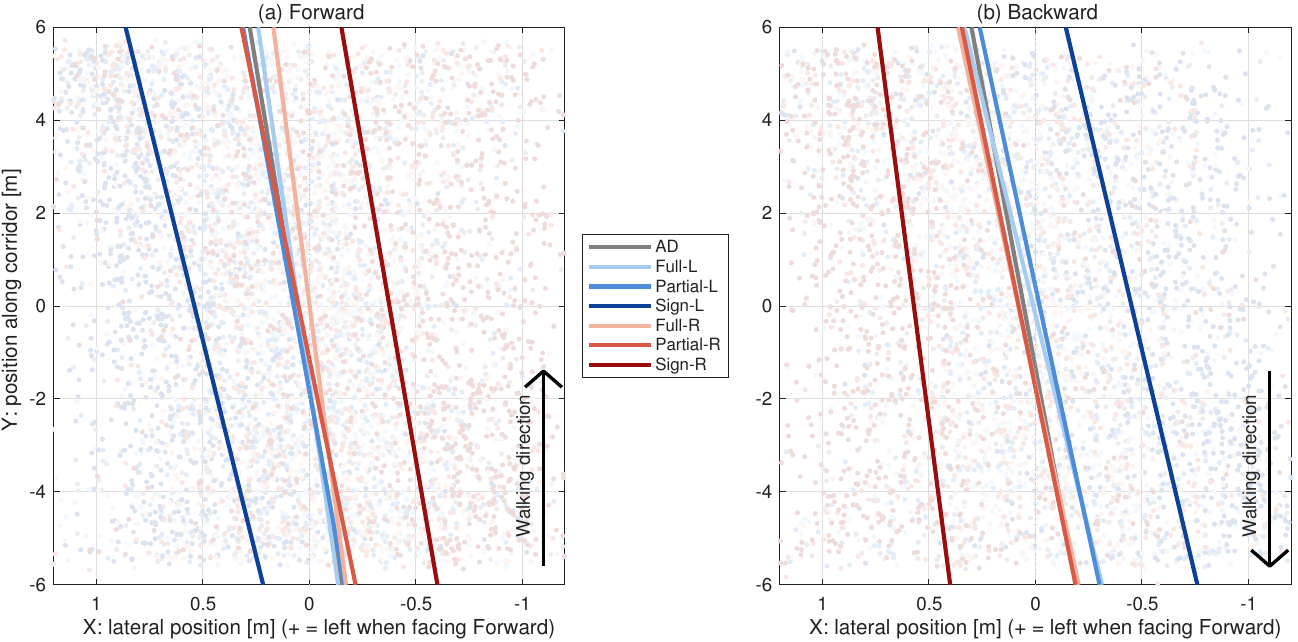}
    \caption{Lateral ($X$) position versus depth ($Y$) in TC2 for the AD, Full, Partial, and Sign conditions. Each point is the mean of frame-wise median centroid positions within a retained 10-s window containing data in at least 10 frames. Colors indicate stimulus direction (Left: blue; Right: red; AD: gray), with shades distinguishing stimulus types. Left and Right refer to the pedestrian's viewpoint, whereas $X$ and $Y$ retain the fixed corridor coordinates; positive $X$ is left when facing Forward. (a) Forward walking. (b) Backward walking. Both panels use the same axis orientations; walking-direction arrows point upward for Forward ($+Y$) and downward for Backward ($-Y$). The inferential position models use $X_{\mathrm{ego}}$, with the Backward $X$ sign reversed. Solid lines are descriptive linear fits for each condition and walking direction, separate from the inferential models. Sign is shown as a descriptive reference.}
    \label{fig:trajectory2}
\end{figure}

\begin{figure}[t]
    \centering
    \includegraphics[width=\linewidth]{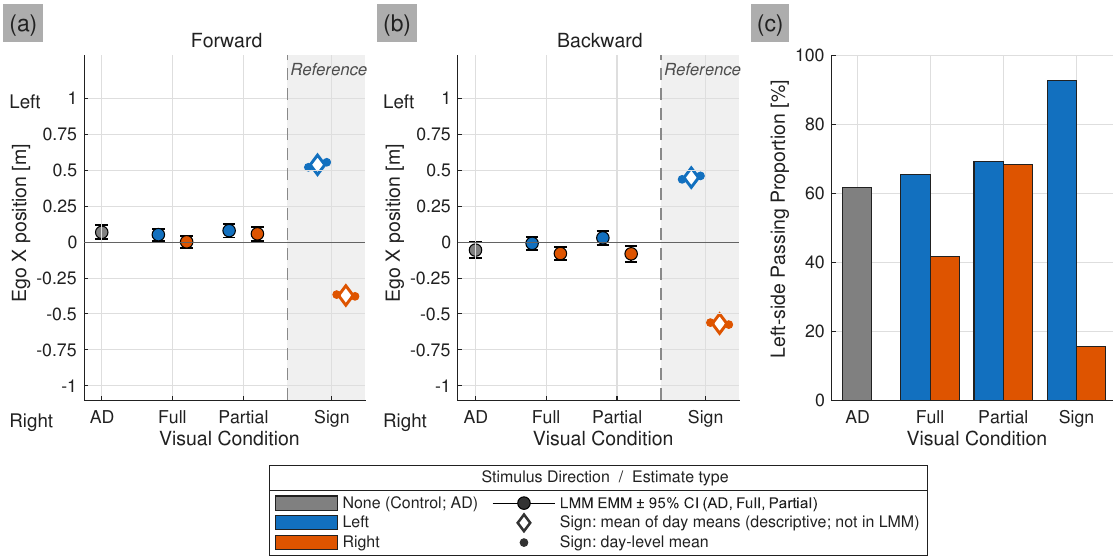}
    \caption{TC2 lateral position and passing-side choice. (a) Forward and (b) Backward lateral positions in walker-referenced coordinates ($X_{\mathrm{ego}}$). Filled circles and error bars for AD, Full, and Partial show estimated marginal means and pointwise, unadjusted 95\% CIs from the five-condition LMM (Sec.~\ref{sec:exp2-oneway-dunnett}), averaged equally over DayType. In contrast, Sign is a descriptive reference excluded from the model (shaded area): small dots show daily means and open diamonds their equally weighted two-day means, without CIs. (c) Observed percentages of left-side passing events, the outcome used for comparison with AD (Sec.~\ref{sec:exp2-passing}). L/R denotes stimulus direction relative to the pedestrian; positive $X_{\mathrm{ego}}$ is to each pedestrian's own left in both walking directions.}
    \label{fig:res2}
\end{figure}

Fig.~\ref{fig:trajectory2} shows lateral position in the fixed corridor frame across depth for each visual condition and walking direction. Fig.~\ref{fig:res2} shows estimated marginal means (EMMs) of $X_{\mathrm{ego}}$ position for AD and the stripe conditions alongside descriptive Sign references (a, b), and observed left-side passing proportions (c).

\subsubsection{Comparison of one-way traffic by stimulus direction, coverage pattern, and walking direction}\label{sec:exp2-oneway-3way}

To examine stripe direction and coverage pattern within each walking direction, we analyzed the 5,942 nonoverlapping 10-s blocks across 10 measurement days from Full-L, Full-R, Partial-L, and Partial-R. Each retained block contained at least 10 frames with valid centroid observations. The outcome was $X_{\mathrm{ego}}$. The LMM included StimDirc (Left and Right), CovPat (coverage pattern; Full and Partial), WalkDirc (Forward and Backward), and DayType (weekday vs. weekend/holiday) as fixed effects, with Day as a random intercept:

\begin{equation}
X_{\mathrm{ego}} \sim \text{StimDirc} \times \text{CovPat} \times \text{WalkDirc} + \text{DayType} + (1\mid\text{Day}).
\end{equation}

Lateral positions differed significantly between Left and Right stripe-motion conditions (StimDirc; $F(1,4.4)=13.09$, $p=.019$, $\eta^2_p=.748$), with an estimated Left--Right difference of 6.4 cm (95\% CI $[1.6,11.1]$ cm), averaged over coverage pattern, walking direction, and day type. WalkDirc was also significant ($F(1,5926.5)=34.36$, $p<.001$, $\eta^2_p=.006$). The WalkDirc effect is consistent with an overall bias toward the Geo-Cosmos, on the left when facing Forward. Neither CovPat ($F(1,4.3)=2.66$, $p=.173$, $\eta^2_p=.382$) nor DayType ($F(1,4.2)=0.76$, $p=.431$, $\eta^2_p=.154$) was significant. None of the interactions was significant (all $p\geq.053$).

In the supplementary analyses of three equal 4-m depth intervals, StimDirc estimates indicated more positive positions under Left than Right in all three zones, but the 95\% confidence interval for this difference excluded zero only in the middle third. Detailed results are provided in the SI (Sec.~\ref*{si-sec:tc2-depth-zones}).

Thus, TC2 showed a directional difference in lateral position between Left and Right stimuli, with the clearest evidence near the center of the walking space. Unlike TC1, no significant difference between coverage patterns was detected.

\subsubsection{Comparison of one-way traffic with the control condition}\label{sec:exp2-oneway-dunnett}

To compare the stripe conditions with the AD control, we analyzed 7,164 retained 10-s blocks across 12 measurement days from AD, Full-L, Full-R, Partial-L, and Partial-R. The outcome was $X_{\mathrm{ego}}$. The LMM included VisualCondition (the five visual conditions), WalkDirc, and DayType as fixed effects, with Day as a random intercept:

\begin{equation}
X_{\mathrm{ego}} \sim \text{VisualCondition} \times \text{WalkDirc} + \text{DayType} + (1\mid\text{Day}).
\end{equation}

WalkDirc had a significant main effect ($F(1,7138.3)=48.46$, $p<.001$, $\eta^2_p=.007$). The WalkDirc effect is consistent with an overall bias toward the Geo-Cosmos, on the left when facing Forward. However, neither VisualCondition ($F(4,5.1)=4.45$, $p=.064$, $\eta^2_p=.777$), DayType ($F(1,5.0)=0.43$, $p=.543$, $\eta^2_p=.079$), nor the VisualCondition$\times$WalkDirc interaction ($F(4,7117.0)=1.69$, $p=.148$, $\eta^2_p<.001$) was significant. No stripe condition differed significantly from AD in either walking direction after Dunnett adjustment (all $p\geq.079$). The corresponding analyses in Front, Middle, and Rear likewise found no significant difference between any stripe condition and AD in either walking direction (SI, Sec.~\ref*{si-sec:tc2-depth-zones}). Thus, these comparisons did not establish a clear positional guidance benefit of moving stripes over advertisement-only signage in TC2.

\subsubsection{Comparison of passing-side choice with the control condition}\label{sec:exp2-passing}\label{sec:passing-side-choice}\label{sec:exp2-passing-2x2}

To test whether passing-side preferences differed between the stripe conditions and the advertisement-only control (AD), we compared their probabilities of left-side passing. The analysis included 617 passing events across 12 days in the AD, Full-L, Full-R, Partial-L, and Partial-R conditions. Because AD has no stripe-motion direction, this analysis compared left-side passing rather than stimulus-congruent passing. Here, $\text{Success}_{\mathrm{left}}=1$ denotes left-side passing and 0 denotes right-side passing for every condition, independently of stimulus direction. We fitted a generalized linear mixed model (GLMM; binomial family, logit link) with VisualCondition and DayType as fixed effects and Day as a random intercept:

\begin{equation}
    \text{Success}_{\mathrm{left}} \sim \text{VisualCondition} + \text{DayType} + (1\mid\text{Day}),
\end{equation}

Likelihood-ratio tests found no significant effect of VisualCondition ($\chi^2(4)=8.52$, $p=.074$) or DayType ($\chi^2(1)=0.006$, $p=.940$). The four stripe--AD comparisons used Dunnett adjustment. The model-estimated probability of left-side passing, averaged over DayType, was numerically lower under Full-R than under AD (0.404 vs. 0.618; difference, $-21.4$ percentage points), but the odds-ratio comparison was not statistically significant ($OR=0.42$, simultaneous 95\% CI $[0.12,1.42]$, $z=-1.74$, $p=.242$, Cohen's $h=-.432$). Full-L, Partial-L, and Partial-R also did not differ significantly from AD (all $p\geq.574$).

An exploratory supplementary analysis of the four stripe conditions detected a significant interaction between stripe-motion direction and coverage pattern, but no overall Full--Partial difference in stimulus-congruent passing. The model-adjusted pooled probability was 56.4\% and did not differ significantly from 50\% (SI, Sec.~\ref*{si-sec:passing-factorial}).

\subsection{Questionnaire response analysis}\label{sec:exp2-questionnaire}

Although iPads were placed at both exits of the MDS space, responses could not be distinguished by exit. The questionnaire analyses therefore combined subjective responses following Forward and Backward walking. The questionnaire response plot, analysis methods, and detailed results accounting for day-level variation are provided in the SI (Fig.~\ref*{si-fig:response2}; Sec.~\ref*{si-sec:questionnaire-tc2}). Among the stripe conditions, Partial received lower Sickness and higher Audiovisual harmony ratings than Full, while Following floor lines ratings showed an interaction between stripe direction and coverage pattern. Compared with AD, Awareness of others was lower under Full-R, Partial-L, and Partial-R, and Sickness was higher under Full-L. The lower Sickness under Partial than Full was consistent with TC1.

\subsection{Summary for Test Case 2}

Although pedestrian lateral positions differed significantly between the leftward- and rightward-motion conditions, none of the individual stripe conditions differed significantly from the advertisement-only control (Secs.~\ref{sec:exp2-oneway-3way} and \ref{sec:exp2-oneway-dunnett}).
No significant Full--Partial positional difference or stripe--AD difference in passing-side choice was detected (Secs.~\ref{sec:exp2-oneway-3way} and \ref{sec:exp2-passing}).
Full received higher Sickness ratings than Partial (Sec.~\ref{sec:exp2-questionnaire}).

These findings leave the stimulus and spatial conditions needed for a practical guidance benefit unresolved, despite Partial's lower discomfort.

\section{Integrated discussion}

The two exploratory field studies examined how pedestrian spatial distributions were associated with visual-motion stimuli under different display layouts and pedestrian flows. We discuss visual motion and pedestrian behavior across spatial contexts, the practical implications of motion pattern design and subjective discomfort, and the descriptive comparison with explicit directional cues.

\subsection{Visual motion and pedestrian behavior across spatial contexts}\label{sec:int-discuss-full-partial}\label{sec:int-discuss-optimal-design}

In TC1, Partial tended to show a greater lateral bias toward stripe motion than Full, although the difference narrowed toward the rear of the display sequence at Large spacing, suggesting the practical potential of subtle visual-motion cues for pedestrian guidance.
In TC2, lateral position differed significantly between the Left and Right stripe-motion conditions, broadly consistent with \citet{mikawa-vection2026}. However, no significant Full--Partial difference emerged in lateral position, and no stripe--AD comparison was significant for lateral position or passing-side choice.

These contrasting patterns raise the question of how pedestrian responses to visual motion vary across spatial contexts. TC1 and TC2 differed in several respects, including display arrangement, advertising content, and pedestrian flow (Table~\ref{tbl:case-characteristics}), making it difficult to isolate the contributions of individual contextual factors. Nevertheless, differences in ambient brightness, visual attention, and lateral display spacing may help explain the observed lateral-position patterns, as previous public-display research suggests that attentional competition \citep{huang2008overcoming,muller2009display,parker2018public} and display arrangement \citep{tenkoppel2012chained} can influence passer-by behavior. In TC2, the displays were located closer to windows and the Geo-Cosmos, an iconic installation at the museum, where brighter surroundings and potential competition for visual attention may have affected the conspicuity of the moving stripes. In addition, lateral display spacing was greater in TC2 (4.5 m) than in TC1 (3.5 m), providing pedestrians with a wider corridor in which to choose their lateral position.

Future studies could extend the display sequence and independently vary ambient lighting, lateral display spacing, and the presence of competing visual stimuli. Pairing these comparisons with field observations, luminance measurements, eye tracking, and ratings of vection intensity could clarify how pedestrians respond to visual motion under different viewing conditions.

\subsection{Practical implications of motion pattern design and subjective discomfort}

Practical motion pattern design needs to consider behavioral responses, subjective discomfort, and compatibility with existing advertising content together. Partial and Circle illustrate how advertising can be accommodated alongside visual motion. The lower Sickness ratings under Partial than Full in both test cases make reduced motion coverage relevant to this balance, although Sickness was a non-focal questionnaire item.

Together with the spatial differences discussed in Sec.~\ref{sec:int-discuss-full-partial}, these findings motivate evaluating designs that accommodate advertising or reduce the salience of visual motion within their intended display layouts.

\subsection{Descriptive comparison with explicit directional cues (Sign)}\label{sec:int-discuss-sign}
Our aim was subtle, non-instructional pedestrian guidance through visual motion alone, without symbols that explicitly indicate a direction. The contrasting Sign conditions in TC2 therefore provide a useful reference. Descriptively, Sign showed more pronounced directional biases in lateral position and passing-side choice than AD or the stripe conditions (Fig.~\ref{fig:res2}). Although visual-motion guidance may be useful only in some settings (Sec.~\ref{sec:int-discuss-full-partial}), stimulus and spatial design should account for the possibility that visual-motion stimuli produce weaker directional responses than explicit instructions.

\section{Limitations}\label{sec:limitations}

In TC1, each condition was measured on a single day, leaving condition differences confounded with day-specific variation in pedestrian volume, composition, and other observed or unobserved factors. Within each spacing, Full differed in DayType from Circle and Partial, whereas Circle and Partial were matched for day type. TC1 also lacked a static or no-motion baseline: the museum considered stationary stripe patterns and static advertisement-only presentations unsuitable for the exhibition, and operational constraints prevented the inclusion of additional control conditions. Consequently, TC1 supports comparisons among moving-stripe conditions without isolating the effect of visual motion itself.

In TC2, measuring each visual condition on multiple days reduced the dependence on any one day, but conditions and day types could not be fully counterbalanced because of space availability, museum closures, and scheduled events. Differences in pedestrian composition and flow across days or times of day may therefore remain. In addition, detected bidirectional encounters were relatively infrequent, ranging from 13 to 285 per day. This low encounter frequency may limit the generalizability of the findings to more crowded settings, such as railway stations.

Despite our efforts to analyze pedestrian flow using LiDAR, the clustering and short-term correspondence estimation used in this study did not allow reliable reconstruction of long-term individual trajectories. A limitation of the position analysis is therefore that the 10-s averages of frame-wise median centroid positions summarize spatial distributions rather than individual-level observations. Because entry to the free-admission area was unrestricted, some visitors may have participated on multiple days; repeat participation could not be identified or accounted for. Temporal aggregation over 10 s reduces short-term dependence but does not guarantee independence between blocks.

Finally, TC1 and TC2 differed simultaneously in several spatial and operational features, including brightness, landmark proximity, display arrangement, pedestrian flow, and advertisement content. The cross-case comparison cannot isolate their causal contributions, and separate within-case significance tests do not test whether effect sizes differ between cases. Luminance, perceived display contrast, gaze allocation, and vection intensity were not directly measured, so the perceptual mechanisms underlying the positional patterns remain unresolved. These limitations constrain generalization beyond the studied museum settings and motivate controlled comparisons alongside further field deployments.

\section{Conclusion}

This exploratory study examined visual-motion-based pedestrian guidance using MDS in two museum field deployments involving 18,402 pedestrian passages over 25 experimental days.
In TC1, the Partial condition tended to show the greatest lateral bias toward stripe motion despite having the smallest stripe coverage. Reducing stripe contrast was not associated with a detectable reduction in the observed lateral bias toward stripe motion. These findings support the practical potential of more subtle, lower-contrast and spatially restricted visual-motion cues for pedestrian guidance.
In TC2, stripe-motion direction was associated with lateral position, but no significant difference was detected between the Full and Partial conditions. None of the stripe conditions differed significantly from the advertisement-only control in either lateral position or passing-side choice. The contrasting patterns in TC1 and TC2 suggest that spatial context may shape how pedestrians respond to stripe coverage. Across both studies, self-reported sickness ratings were lower with partial stripes than with full-screen stripes.

Together, these findings suggest that lower-contrast and spatially restricted visual-motion cues hold promise for pedestrian guidance in some spatial contexts and that spatially restricted guidance cues may coexist with existing advertising while reducing discomfort. 
Our results highlight the importance of considering stimulus contrast, motion coverage, spatial configuration, and pedestrian comfort together when designing visual-motion-based guidance systems for public spaces.

\backmatter
\bmhead{Acknowledgements}
We sincerely thank the National Museum of Emerging Science and Innovation (Miraikan) in Japan for providing the experimental venues and for their generous organizational and operational support throughout the study.

\section*{Declarations}
\begin{itemize}
\item Funding: This work was supported by JST PRESTO Grant Number JPMJPR25IA.
\item Conflict of interest/Competing interests:
Not applicable.

\item Ethics approval and consent to participate:
This study was reviewed and approved by the Ethics Committee of NTT Communication Science Laboratories as protocol R07-013 on August 18, 2025. For the public experiment, individual written informed consent forms were not collected; instead, an ethics-approved notice-based opt-out consent procedure was used, including prior notification at the venue, staff guidance to an adjacent avoidance route for non-participants, and instructions to notify staff for opting out of point-cloud data collection.

\item Consent for publication: Not applicable. %
\item Data availability:
The full datasets, including the raw LiDAR point-cloud data, are not available because of their large size ($\geq$2.0 TB). However, the point-cloud data cropped to the interior MDS spaces, together with the questionnaire responses, do not contain personally identifiable information and are available upon reasonable request. Requests should be directed to YM at yuri\_mikawa@ipc.i.u-tokyo.ac.jp.

\item Materials availability: Not applicable. %
\item Code availability: The analysis code is available from the corresponding author upon reasonable request.
\item Author contributions:
YM: Conceptualization, Data curation, Formal analysis, Funding acquisition, Investigation, Supervision, Visualization, Writing – original draft, Writing – review \& editing.\\
TF: Investigation, Methodology, Resources, Writing – review \& editing.\\
YK: Investigation, Methodology, Writing – review \& editing.\\
TY: Investigation, Methodology, Writing – review \& editing.\\
KM: Investigation, Project administration, Writing – review \& editing.

\item Generative AI and AI-assisted technologies: Generative AI tools were used for English-language editing and assistance in drafting and checking analysis code. All analyses and outputs were executed, inspected, and verified by the authors, who take full responsibility for the reported methods, results, and conclusions. %
\end{itemize}

\clearpage

\clearpage
\bibliography{citation}

\end{document}